# Complexity analysis of the turbulent environmental fluid flow time series

**D.T. Mihailović[a,*], E. Nikolić-Đorić[b], N. Drešković[c] and G. Mimić[d]**

[a] Faculty of Agriculture, Division of Meteorology, University of Novi Sad, Dositeja Obradovica Sq. 8, 21000 Novi Sad, Serbia

[b] Faculty of Agriculture, Division of Statistics, University of Novi Sad, Dositeja Obradovica Sq. 8, 21000 Novi Sad, Serbia

[c] Faculty of Sciences, Department of Geography, University of Sarajevo, Zmaj from Bosnia 33-35, 71000 Sarajevo, Bosnia and Herzegovina

[d] Faculty of Sciences, Department of Physics, University of Novi Sad, Dositeja Obradovica Sq. 3, 21000 Novi Sad, Serbia

**Abstract**

We have used the Kolmogorov complexities, sample and permutation entropies to quantify the randomness degree in river flow time series of two mountain rivers in Bosnia and Herzegovina, representing the turbulent environmental fluid, for the period 1926–1990. In particular, we have examined the monthly river flow time series from two rivers (Miljacka and Bosnia) in mountain part of their flow and then calculated the Kolmogorov Complexity (KL) based on the Lempel–Ziv Algorithm (LZA) (Lower – KLL and Upper - KLU), Sample Entropy (SE) and Permutation Entropy (PE) values for each time series. The results indicate that the KLL, KLU, SE and PE values in two rivers are close to each other regardless of the amplitude differences in their monthly flow rates. We have illustrated the changes in mountain river flow complexity by experiments using (i) the data set for the Bosnia River and (ii) anticipated human activities and projected climate changes. We have explored the sensitivity of considered measures in dependence on the length of time series. In addition, we have divided the period 1926–1990 into three subintervals: (a) 1926 -1945, (b) 1946–1965, (c) 1966–1990, and calculated the KLL, KLU, SE, PE values for the various time series in these subintervals. It is found that during the period 1946 - 1965, there is a decrease in their complexities, and corresponding changes in the SE and PE, in comparison to the period 1926–1990. This complexity loss may be primarily attributed to (i) human interventions, after the Second World War, on these two rivers because of their use for water consumption and (ii) climate change in recent time.

---
[*] Corresponding author. Tel. +381216350552.

*E-mail address*: guto@polj.uns.ac.rs (D.T. Mihailović)

# 1 Introduction

Scientists in different fields (physicists, meteorologists, geologists, hydrologists, engineers etc., among others) study environmental fluid motion. Behavior of these fluids are significantly influenced by (i) human activities, (ii) climatic change and (iii) increasing water pollution changing mass and energy balance of the fluid. Understanding their complexity can help us to learn how to improve our systems by understanding how complexity underlies and affects the environments and the systems. Influenced by the aforementioned factors, the river flow in different geographic region may range from being simple to complex, varying in both time and space. For turbulent environmental fluids like mountain rivers the speed of the water flow can vary within a system and is subject to chaotic turbulence. This turbulence results in divergences of flow from the mean downslope flow vector as typified by eddy currents. The mean flow rate vector is based on variability of friction with the bottom or sides of the channel, sinuosity, obstructions, and the incline gradient [1]. Over the last decade controversial results have been obtained about the hypothetical chaotic nature of river flow dynamics [2-5]. For example, Zunino et al. [5] analyzed the streamflow data corresponding to the Grand River at Lansing (Michigan) trying to provide new insights regarding this issue, while Hajian and Sadegh Movahed [6] have used the detrended cross-correlation analysis in order to investigate the influence of sun activity represented by sunspot numbers on river flow fluctuation as one of the climate indicators. The river flow fluctuations also have been analysed using the formalism of the fractal analysis [7]. Therefore, it is of interest to determine the nature of complexity in mountain river flow processes that can not be done by traditional mathematical statistics what requires the use of different measures of complexity. These measures help us to get an insight into the complexity of the environmental fluid flow; i.e. the mountain river flow in this paper. Using them, we can more comprehensively investigate possible changes in: (i) river flow due to human activities, (ii) response to climate changes, (iii) nonlinear dynamic concepts for a catchments classification framework. Also, we will be able to improve application of the stochastic process concept in hydrology for its modeling, forecasting, and other ancillary purposes [2, 8-10].

Kolmogorov Complexity is used in order to describe the complexity or degree of randomness of a binary string. It is in the literature also known as algorithmic entropy, stochastic complexity, descriptive complexity, Kolmogorov-Chaitin complexity and program-size complexity. This measure was independently developed by Andrey N. Kolmogorov in the late 1960s [11]. Later following Kolmogorov's idea, Lempel and Ziv [12] developed an algorithm for calculating the measure of complexity. We will refer to the Lempel-Ziv Algorithm by LZA. It can be considered as a measure of the degree of disorder or irregularity in a time series. This algorithm has been used for evaluation of the randomness present in time series. Entropy is commonly used to characterize the complexity of a time series also including hydrological ones [13,14]. Thus, approximate entropy with a biased statistic, is effective for analyzing the complexity of noisy, medium-sized time series [15]. Richman and Moorman [16] proposed another statistic, sample entropy (SE), which is unbiased and less dependent on data.

Traditional entropies quantify only the regularity of time series having some disadvantages [17]. Permutation entropy (PE), introduced by Bandt and Pompe [18], is a measure based on comparison of neighboring values of time series. The advantage of this measure is its applicability to real data, its robustness if observational noise is present and invariance to non-linear transformations. The different measures implemented in this paper are useful to quantify the degree of randomness present in time series. Also, they may detect structural changes over time. However, they are not able to discriminate the degree of structure present in a process as statistical or structural complexity measures do [19-23].

The SE is not often used in complexity analysis of the environmental fluid flow dynamics, while to our knowledge, in paper by Zunino et al. [5], the PE quantifier was the first time implemented for studying the dynamics of a river's flow.

The purpose of this paper is to consider the complexity of the river flow dynamics of two mountain rivers in Bosnia and Herzegovina for the period 1926–1990, using the KLL, KLU, SE and PE measures. That will be done through: (i) introducing the KLU complexity, (ii) sensitivity tests for all concidered measures in dependance on data length and (iii) their application on two river flow time series.

## 2 Method

*2.1. Description of the LZA algorithm for computing the KL complexity*

The Kolmogorov complexity analysis of a time series $\{x_i\}$, $i = 1, 2, 3, 4, ..., N$ can be carried out as follows. *Step 1:* Encode the time series by constructing a sequence $S$ of the characters 0 and 1 written as $\{s(i)\}$, $i=1,2,3,4,…,N$, according to the rule

$$s(i) = \begin{cases} 0 & x_i < x_* \\ 1 & x_i \geq x_* \end{cases}. \qquad (1)$$

Here $x_*$ is a chosen threshold. We use the mean value of the time series to be the threshold. The mean value of the time series has often been used as the threshold [24]. Depending on the application, other encoding schemes are also used [25-26].

*Step 2:* Calculate the complexity counter $c(N)$. The $c(N)$ is defined as the minimum number of distinct patterns contained in a given character sequence [27]. The complexity counter $c(N)$ is a function of the length of the sequence $N$. The value of $c(N)$ is approaching an ultimate value $b(N)$ as $N$ approaching infinite, i.e.

$$c(N) = O(b(N)), \quad b(N) = \frac{N}{\log_2 N}. \qquad (2)$$

*Step 3*: Calculate the normalized complexity measure $C_k(N)$, which is defined as

$$C_k(N) = \frac{c(N)}{b(N)} = c(N) \frac{\log_2 N}{N}. \qquad (3)$$

The $C_k(N)$ is a parameter to represent the information quantity contained in a time series, and it is to be a 0 for a periodic or regular time series and to be a 1 for a random time series, if $N$ is large enough. For a non-linear time series, $C_k(N)$ is to be between 0 and 1.

Above steps are incorporated in codes of different programming languages to estimate the lower version of the Kolmogorov complexity (KLL). This version is commonly used by researchers. However, there exists the upper version of the Kolmogorov complexity (KLU), which is described in [12]. Note, that in both cases, an extension to a sequence is considered "innovative" in some way, but differently. Here we describe both of them. The LZA is an algorithm, which calculates the KL measure of binary sequence complexity. As inputs it uses a vector $S$ consisting of a binary sequence whose complexity we want to analyze and calculate converting the numeric values to logical values depending on whether (0) or not (1). In this algorithm we can evaluate as a string two types of complexities, which one is "exhaustive", i.e., when complexity measurement is based on decomposing $S$ into an exhaustive production process. On the other hand so called "primitive" complexity measurement is based on decomposing $S$ into a primitive production process. Exhaustive complexity can be considered a lower limit of the complexity measurement approach (KLL) and primitive complexity an upper limit (KLU). Let us note that the "exhaustive" is considered as the KL measure and frequently used in complexity analysis. The KLL calculation is based on finding extensions to a sequence, which are not reproducible from that sequence, using a recursive symbol-copying procedure. The KLU calculation uses the eigenfunction of a sequence. The sequence decomposition occurs at points where the eigenfunction increases in value from the previous one. In this case, the locations where an extra symbol that causes an increase in the accumulated vocabulary.

First, we have to find array consisting of the history components $O$ that were found in the sequence $S$, whilst calculating the KLL or KLU ($C_k$ in (3)). Each element in $O$ consists of a vector of logical values (true, false), and represents a history component. Histories are composed by decomposing the sequence $S$ into the following sequence of words

$$O(S) = S(1,h_1)S(h_1+1,h_2)S(h_2+1,h_3)...S(h_{m-1}+1,h_m) \ , \tag{4}$$

where the indices $\{h_1,h_2,h_3....h_{m-1},h_m\}$ characterise a history making up the set of "terminals". We do not know how long the histories will be or in other words how many terminals we need. As a result, we will allocate an array of length equal to the eigenfunction vector length ($Es(h)$).

For an exhaustive history (i.e. when we calculate the KLL), from Theorem 8 in [12] the terminal points $h_i$, $1 \leq i \leq m-1$, are defined by

$$h_i = min\{h | Es(h) > h_{m-1}\} \ . \tag{5}$$

From the same theorem, for a primitive history (i.e. when we calculate the KLU), the terminal points $h_i$, $1 \leq i \leq m-1$ are defined by

$$h_i = min\{h | Es(h) > Es(h_{i-1})\} \ , \tag{6}$$

where the eigenfunction, $Es(n)$, is monotonically non-decreasing (the Lemma 4 in [12]). Finally, we use the terminal points to get calculate $c_{pr}$ (primitive, KLU) or $c_{ex}$ (exhaustive, KLL), as the length of the production histories $O_{pr}(S)$ or $O_{ex}(S)$, which are so called un-normalized complexities (Eq. (2)). To get normalized ones we use Eq. (3). In this paper we have designed our own code in FORTRAN90, which partly relies on the MATLAB by Thai [28].

*2.2. Calculation of sample entropy*

This is a measure quantifying regularity and complexity; it is believed to be an effective analysing method of diverse settings that include both deterministic chaotic and stochastic processes, particularly operative in the analysis of physiological, sound, climate and environmental interface signals that involve relatively small amount of data [16, 29-30]. The threshold factor or filter $r$ is an important parameter. In principle, with an infinite amount of data, it should approach zero. With finite amounts of data, or with measurement noise, $r$ value typically varies between 10 and 20 percent of the time series standard deviation [31]. To calculate it from a time series, $X = (x_1, x_2, ..., x_N)$, one should follow these steps [16]:

(1) Form a set of vectors $X_1^m, X_2^m, ..., X_{N-m+1}^m$ defined by $X_i^m = (x_i, x_{i+1}, ..., x_{i+m-1})$, $i = 1, ..., N-m+1$;

(2) The distance between $X_i^m$ and $X_j^m$, $d[X_i^m, X_j^m]$ is the maximum absolute difference between their respective scalar components: $d[X_i^m, X_j^m] = \max_{k \in [0, m-1]} |x_{i+k} - x_{j+k}|$;

(3) For a given $X_i^m$, count the number of $j$ ($1 \leq j \leq N-m, j \neq i$), denoted as $B_i$, such that $d[X_i^m, X_j^m] \leq r$. Then, for $1 \leq i \leq N-m$, $B_i^m(r) = B_i/(N-m-1)$;

(4) Define $B^m(r)$ as: $B^m(r) = \{\sum_{i=1}^{N-m} B_i^m(r)\}/(N-m)$;

(5) Similarly, calculate $A_i^m(r)$ as $1/(N-m-1)$ times the number of $j$ ($1 \leq j \leq N-m, j \neq i$), such that the distance between $X_j^{m+1}$ and $X_i^{m+1}$ is less than or equal to $r$. Set $A^m(r)$ as: $A^m(r) = \{\sum_{i=1}^{N-m} A_i^m(r)\}/(N-m)$. Thus, $B^m(r)$ is the probability that two sequences will match for $m$ points, whereas $A^m(r)$ is the probability that two sequences will match $m+1$ points;

(6) Finally, define: $SampEn(m,r) = \lim_{N \to \infty} \{-\ln[A^m(r)/B^m(r)]\}$ which is estimated by the statistic:

$$SampEn(m, r, N) = -\ln \frac{A^m(r)}{B^m(r)}.$$

*2.3. Calculation of permutation entropy*

Permutation entropy, introduced by Bandt and Pompe [18], is the complexity measure based on comparison of neighboring values of time series. The advantage of this measure is its applicability to real data, its robustness if observational noise is present and invariance to non-linear transformations. For $N$ sample time series $\{x(i) : 1 \leq i \leq N\}$, all $m!$ permutations $\pi$ of order $m$ ($m < N$) are considered. The relative frequency for each permutation $\pi$ is

$$p(\pi) = \frac{\#\{i \mid 0 \leq i \leq N-m, (x_{i+1},...,x_{i+m}) \text{ is of type } \pi\}}{N-m+1} \quad . \tag{7}$$

When the underlying stochastic process satisfies a very weak stationary condition that $x_i < x_{i+k}$ for $k \leq m$ is independent of $i$, the relative frequency $p(\pi)$ converges to exact probability if $N \to \infty$.

The permutation entropy of order $m \geq 2$ is defined as $H(m) = \sum_{i=1}^{m!} p(\pi_i) \log p(\pi_i)$. The value of $H(m)$ is always $0 \leq H(m) \leq \log(m!)$ where lower bound is attained for monotone time series (increasing or decreasing), and the upper bound for an identically independent random sequences, when all possible permutations have the same probability. In the experiment with chaotic time series, Bandt and Pompe [16] established that for chaotic time series, $H(m)$ increases almost linearly with $m$.

## 3 Data and computations

*3.1 Short description of river locations and time series*

The River Bosnia and the River Miljacka flow through the Sarajevo Valley, which is located between mountain depressions and between the massive Bjelasnica and Igman mountains on the southwest as well as the low mountains and middle mountains on the northeast. The valley generally stretches in the NW-SE direction and there are low mountains and middle mountain areas on the southeastern slopes of Trebevic Mountain and on the northwestern slopes between valley peaks (Fig. 1).

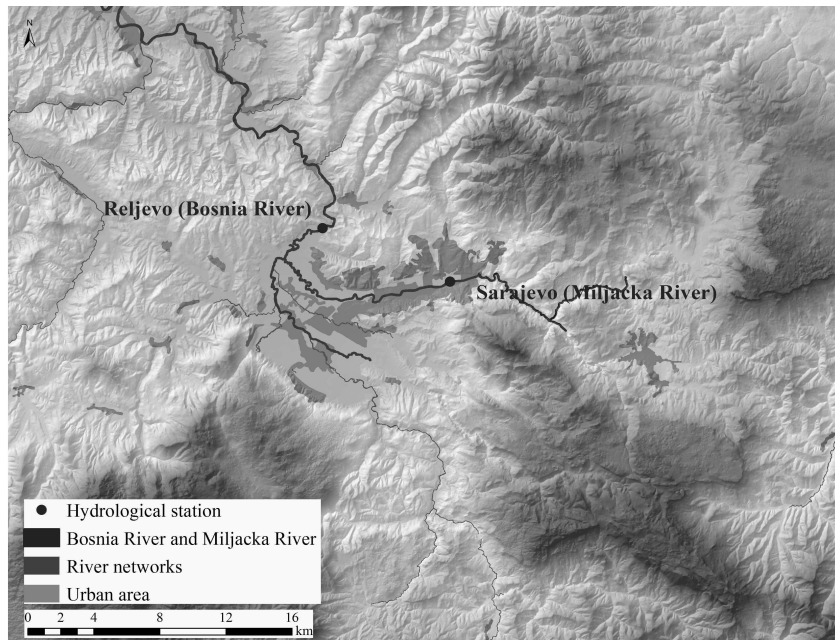

**Fig. 1** Topological location of the Sarajevo Valley with hydrological stations Reljevo (the Bosnia River) and Sarajevo (the Miljacka River) used in this study (designed by N. Drešković).

The mean altitude of the bottom of the valley is approximately 515 m. The valley is a hydrological input for the source area of the Bosnia River with seven tributaries including the Miljacka River. In this part of their flow both of them fully represent mountain rivers. For this study for time series we used monthly mean values (Fig. 2) from hydrological stations Reljevo (the Bosnia River) and Sarajevo (the Miljacka River) since they have representative and reliable instrument for hydrological monitoring since 1926 [32].

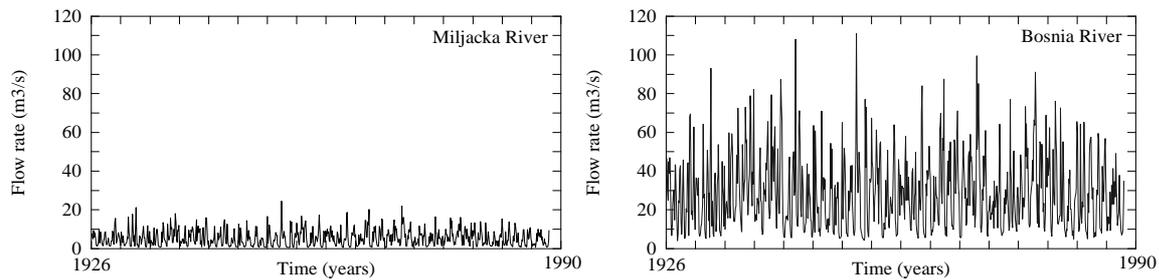

**Fig. 2** River flow time series for the Miljacka River and the Bosnia River for the period 1926-1990.

The Bosnia River has the mean annual river flow about 8.0 $m^3 s^{-1}$, except during the precipitation season when it takes value of 24.0 $m^3 s^{-1}$. The hydrological station Reljevo is located 11.6 km away from its source. Usually the mean annual river flow of this river is 28.7 $m^3 s^{-1}$, with a maximum of 44.9 $m^3 s^{-1}$ (in 1937) and a minimum value of 17.9 $m^3 s^{-1}$ (in 1990) during the period 1926-1990. The entire Miljacka River system upstream has a very steep and wavy longitudinal profile. Downstream from this site, it flows through the alluvial plateau with a very small drop (3 % - 5 %) passing the highly urbanized Sarajevo Valley with over 400,000 inhabitants. The hydrological station Sarajevo is located on the bridge in the central part of Sarajevo. Usually the mean annual river flow of the Miljacka River is 5.5 $m^3 s^{-1}$, with a maximum of 9.1 $m^3 s^{-1}$ (in 1937) and a minimum value of 3.0 $m^3 s^{-1}$ (in 1990) during the period indicated. The river flow time series for the Miljacka River and the Bosnia River for the period 1926–1990 are depicted in Fig. 2.

*3.2. An example of changes in complexity of the turbulent environmental fluid time series*

The mountain river is a typical example of the turbulent environmental fluid for which the changes in complexity of its flow rate depends on human activities and climate change. These process and phenomena can contribute to the loss of the complexity, which leads to reducing the stochastic component in the river flow. Undoubtedly, the nature of its complexity can not be explored by traditional methods of mathematical statistics. Therefore, it requires the use of various measures of complexity to get an insight into the complexity of its flow rate. In an example, that follows, we will illustrate the impact of these mentioned factors on mountain river flow complexity. In these experiments we use the time series for the Bosnia River (the right panel in Fig. 2) to simulate loss of flow complexity of this river as result of the anticipated (i) human activities and (ii) projected climate changes in the region from Fig. 1.

The influence of the human activity (for example, urbanization and building capacities for the water consumption, etc.) on the mountain river flow complexity we have simulated artificially. Namely, when a value of the KL is close zero then it is associated with a simple deterministic process like a periodic motion, whereas a value close to one is associated with a stochastic process [32]. Thus, by human activities, from the flow of the turbulent river many stochastic components can disappear in dependence on the level and intensity of those activities. The influence of the human activity on the river flow complexity we have simulated in the following way. First, depending on the intensity of activity (symbolically depicted in percentage on $x$ axis in Fig. 3b): (i) we have removed amplitudes in the time series setting them to be zero and (ii) we have kept those samples in the time series always having the size $N$ ($N = 780$ in this experiment). Then, using the procedure described in subsection 2.1 we have calculated the KLL for each created time series.

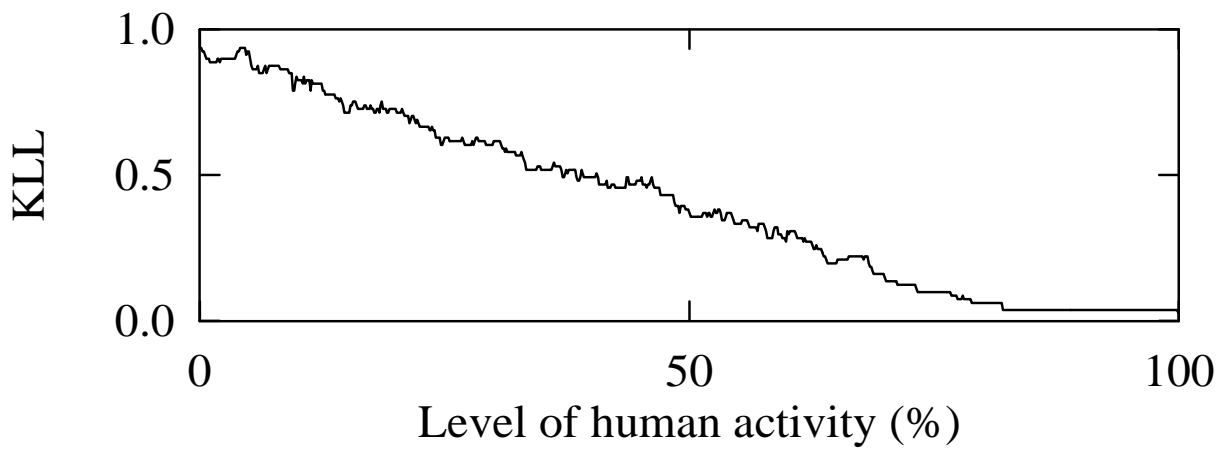

**Fig. 3** Changes in the KLL complexity of a mountain river flow in dependence on the level of human activity.

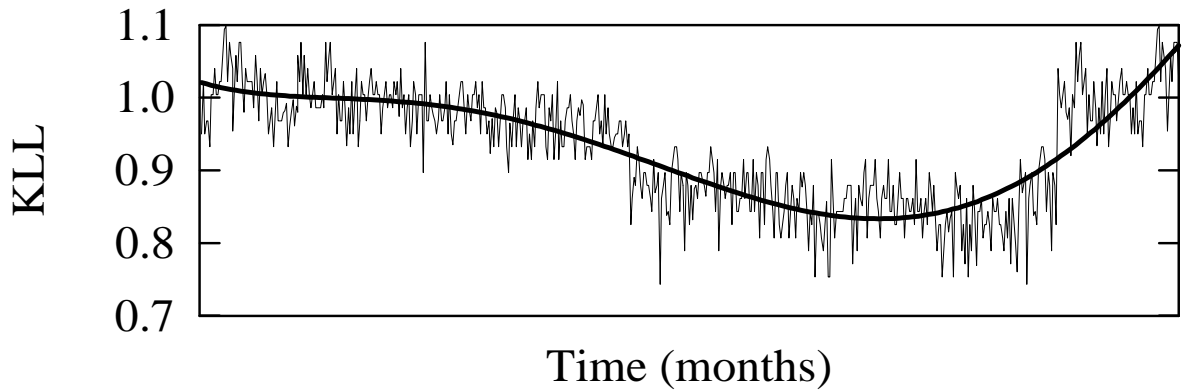

**Fig. 4** Changes in the KLL complexity of a mountain river flow rate in dependence on simulated climate changes. Heavy solid line is a fitting curve, which depicts the trend of the complexity change.

Changes in the KLL complexity of a mountain river flow rate in dependence on simulated human activity are depicted in Fig. 3b. From this figure it is seen a descending trend of this curve, which is finished by a straight line on the lowest level of complexity depicting the absence of turbulent eddies as a result of regularization of the river flow. The descending curve is rather wavy then linear because of the nonlinearity of the river flow.

The climate change impact on the mountain river flow complexity has been simulated artificially in the following way. The time series of the flow rate was divided into three subintervals: (1) 1- $280^{th}$, (2) 280 - $520^{th}$ and (3) 580 - $780^{th}$ month. The impact of climate change on the river flow complexity was introduced during the period (2) of simulation following the results of regional climate simulations by Djurdjevic and Rajkovic [34] that includes area depicted in Fig. 1. According to them, projections for 2030 year indicate an evident increase of air temperature and evaporation (about 20 %) as well as the decrease of precipitation. For the periods (1) and (3) we have calculated the KLL in subsection 2.1. For the period (2), first we have recalculated the monthly river flow rates by changing their values, according to values of evaporation and precipitation obtained by the regional climate model [34], and then we have applied the same procedure for the KLL calculations as in previous experiment. As a consequence of that, from Fig. 4 it is seen an evident decrease of the complexity of the river flow time series, which is visualized through the fitting curve

*3.3 Computation of measures for two river flow time series*

Using the calculation procedure outlined in subsections 2.1-2.3, we have computed the KLL, KLU, SE and PE values for the two river flow time series. The calculations are carried out for the entire time interval 1926–1990 and for three subintervals covering this period: (a) 1926–1945, (b) 1946–1965 and (c) 1966–1990 obtained by sensitivity tests in dependence on length of time series.

*3.4 Sensitivity tests*

According to previous results all measures are sensitive to the length of time series, *N*. For the SE, there exists a recommendation for use *N* that is larger than 200 [35]. For the PE the length of the time series must be larger than the factorial of the embedding dimension [36]. Let us note that Hu et al. [37] derived analytic expression for $C_k$ (notation in subsection 2.1) in the KLL, for regular and random sequences. In addition they showed that the shorter length of the time series, the larger $C_k$ value and correspondingly the complexity for a random sequence can be considerably larger than 1. In order to explore the sensitivity of these measures in dependence on the length of time series we calculated the KLL, KLU, SE and PE values for $N=200$ up to $N=780$ (Fig. 5). In these experiments we have had in mind the following facts. The SE is sensitive on input parameters: embedding dimension ($m$), tolerance ($r$) and time delay ($\tau$). In this study it was calculated for river flow time series with the following values of parameters: $m=2$, $r=0.2$ and $\tau=1$. Beside $N$, the embedding dimension ($m$), also called as the permutation order, is an input parameter for PE. Therefore we have considered its sensitivity on the PE outputs. Due to the length of time series ($N=780$) we chose the embedding dimension to be less then 6 (Fig. 6).

Our results indicate that the KLL and SE decrease and the KLU and PE slightly increase when the number of observations increases. All considered measures are sensitive to random component and may be considered as indicators of randomness, but they do not give information about amplitude variations. In particular, we have calculated the frequencies of the river flow time series. They have the same dominant frequencies (1/12 and 1/6 for the Miljacka River and the Bosnia River, respectively) as well as the similar distribution of the random component. Thus the values of complexities, calculated for the whole time series and subintervals for both rivers, are close to each other.

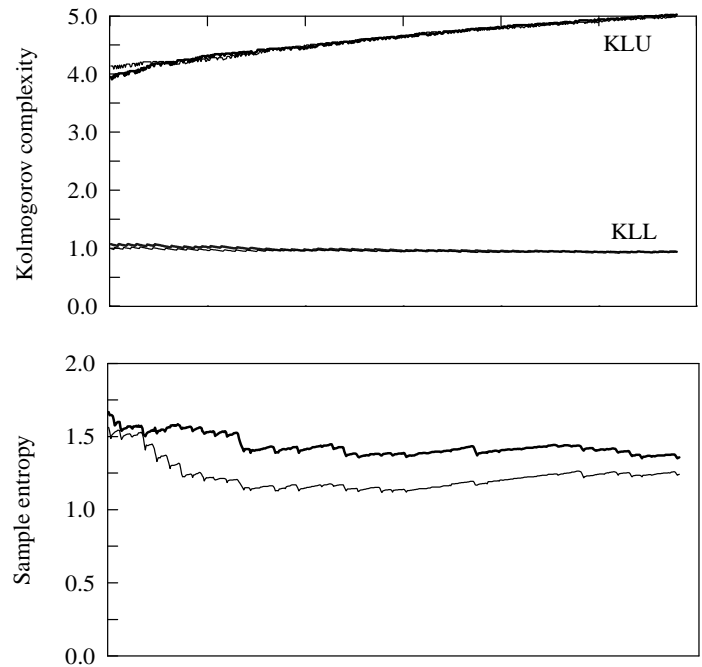

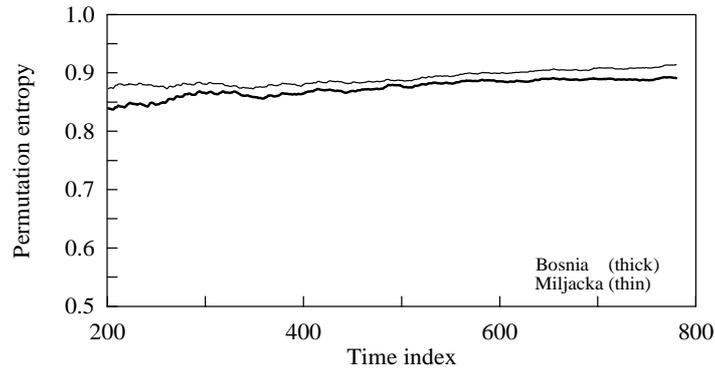

**Fig. 5** Sensitivity of the KLL, KLU (upper), SE (middle) PE (lower) panel in dependence on the length of the river flow time series for the Miljacka River and the Bosnia River.

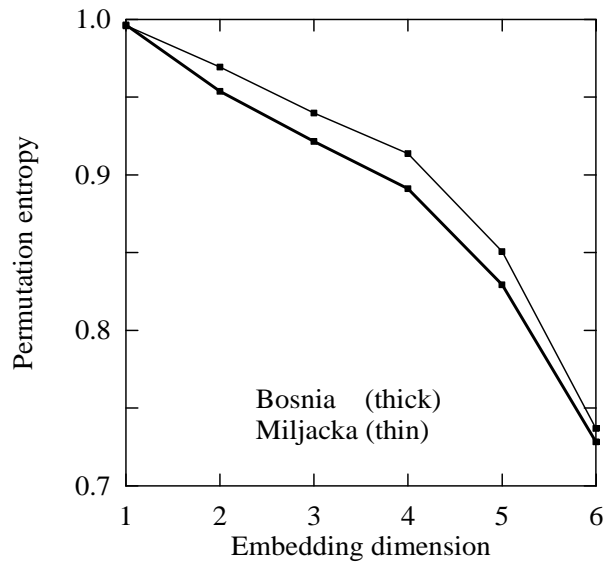

**Fig. 6** Permutation entropy as a function of embedding dimension for river flow time series for the Miljacka River and the Bosnia River for the period 1926-1990.

## 4 Results and comments

Using the calculation procedure presented in subsections 2.1-2.3, we have computed the KLL, KLU, SE and PE values for river flow time series of two rivers. The calculations are carried out for the entire time interval 1926–1990. The results are given in the corresponding rows of Table 1. It is seen from this table that the KLL values in both rivers are close while the KLU ones practically the same. Note that a process that is least complex has a KLL value near to zero, whereas a process with highest complexity will have KLL close to one. The KLL measure can be also considered as a measure of randomness. Thus, a value of the KLL near zero is associated with a simple deterministic process like a periodic motion, whereas a value close to one is associated with a stochastic process [33]. Accordingly, the KLL values, which are large for both rivers (0.936), point out the presence of stochastic influence in these typically mountain rivers. The other two calculated measures indicate on a similar behavior of time series for both rivers, i.e. their increased irregularity. The SE values are slightly different (1.240 for Mil and 1.357 for Bos) while the PE values are very close to each other (0.914 for Mil and 0.891 for Bos).

| River | Measure | 1926-1990 | 1926-1945 | 1946-1965 | 1966-1990 |
|---|---|---|---|---|---|
| Miljacka | KLL | 0.936 | 0.988 | 0.955 | 0.988 |
| (Mil) | KLU | 5.002 | 4.210 | 3.944 | 4.557 |
|  | SE | 1.240 | 1.438 | 0.903 | 1.478 |
|  | PE | 0.914 | 0.879 | 0.832 | 0.903 |
|  |  |  |  |  |  |
| Bosnia | KLL | 0.936 | 1.054 | 0.977 | 0.988 |
| (Bos) | KLU | 5.024 | 4.103 | 4.031 | 4.471 |
|  | SE | 1.357 | 1.526 | 1.214 | 1.367 |
|  | PE | 0.891 | 0.843 | 0.847 | 0.869 |

**Table 1** Kolmogorov complexities (lower – KLL and upper - KLU), sample entropy (SE) and permutation entropy (PE) values for the river flow time series of two mountain rivers for the period 1926–1990, and the subintervals: (a) 1926–1945, (b) 1946–1965, (c) 1966–1990. In computing the entropies we have used the following sets of parameters ($m=2$, $r=0.2$ and $\tau=1$) and ($m=5$) for the SE and PE, respectively.

We have also divided the period 1926–1990 into three subintervals: (a) 1926–1945, (b) 1946–1965, (c) 1966–1990, and calculated the KLL, SE and PE values for the various time series in each of these subintervals. These intervals were chosen from two reasons. Firstly, it was expected a change in the complexity of both rivers in the period 1945 (end of the Second World War) - 1965 (end of the most intensive human intervention, in particular, urbanization and building capacities for the water consumption). Let us note complexity in river flow time series may be lost due to the different human activities [38-39]. Secondly, we have performed the sensitivity tests (subsection 2.3) to check reliability of chosen time series of subintervals. On basis those tests, in computing procedure we have used the following parameters: (ii) embedding dimension ($m=2$), tolerance ($r=0.2$) and time delay ($\tau=1$) for the SE and (ii) embedding dimension ($m=5$) for the PE. In result the time series for periods (a), (b) and (c) were 240, 240 and 300, respectively.

It is found that during 1946–1965, there is a decrease in complexity in Mil and Bos rivers (0.955 and 0.977, respectively) in comparison to the other subintervals. This complexity loss may be interpreted as results of intensive different human activities on those rivers after the Second World War. The same result is found for the KLU complexity, i.e., 3.944 for Mil and 4.031 for Bos, what are the lowest their values in comparison to the other subintervals. Lower values of both entropies for both rivers: (i) the SE (Mil-0.903; Bos-1.214) and (ii) the PE (Mil-0.832), support conclusion about more regular river flow time series in this period.
Only in the case of PE there is minor decrement of the regularity for the period 1946-1965. In the case of the PE, the same conclusion holds for other considered values of embedding dimension.

## 5   Concluding remarks

In the present study we have analyzed monthly river flow to assess the complexity in river flow dynamics of two rivers in Bosnia and Herzegovina (Miljacka and Bosnia) for the period 1926–1990. We have examined the monthly river flow time series from two rivers (Miljacka and Bosnia) in the mountain part of their flow and calculated the KLL, KLU, SE and PE values for each time series. We have illustrated the changes in mountain river flow complexity by simulation experiments using (i) the data set for the Bosnia River and (ii) anticipated human activities and projected climate changes in the region. We have performed sensitivity tests with the lengths of the time series to choose reliable length for subintervals in which we divided the entire time series. According to all computed measures, except PE for River Bosnia, it is found that during 1946–1965, there is a decrease in complexity in the River Miljacka and the River Bosnia in comparison to the other chosen subintervals. This complexity loss may be interpreted as results of intensive different human intervention on those rivers after the Second World War.

**Acknowledgements**

This paper was realized as a part of the project "Studying climate change and its influence on the environment: impacts, adaptation and mitigation" (43007) financed by the Ministry of Education and Science of the Republic of Serbia within the framework of integrated and interdisciplinary research for the period 2011-2014.

**References**

[1] J.D. Allan, Stream Ecology: structure and function of running waters, Chapman and Hall, London, 1995, pp. 388.

[2] D. Schertzer, I. Tchiguirinskaia, S. Lovejoy, P. Hubert, H. Bendjoudi, and M. Larchevesque, DISCUSSION of „Evidence of chaos in rainfall-runoff process" Which chaos in rainfall-runoff process?, Hydrol. Sci. J. 47 (2002) 139-149.

[3] B. Sivakumar, V.P. Singh, Hydrologic system complexity and nonlinear dynamic concepts for a catchment classification framework, Hydrol. Earth Syst. Sc. 16 (2012) 4119–4131.

[4] J.D. Salas, H.S. Kim, R. Eykholt, P. Burlando and T.R. Green, Aggregation and sampling in deterministic chaos: implications for chaos identification in hydrological processes, Nonlinear Proc. Geoph. 12(4) (2005) 557-567.

[5] L. Zunino, M. C. Soriano, O. A. Rosso, Distinguishing chaotic and stochastic dynamics from time series by using a multiscale symbolic approach Phys. Rev. E 86, 046210 (2012).

[6] S. Hajian, M. Sadegh Movahed, Multifractal Detrended Cross- Correlation Analysis of sunspot numbers and river flow fluctuations Physica A 389 (2010) 4942-4957.

[7] M. Sadegh Movahed and E. Hermanis, Fractal Analysis of River Flow Fluctuations, Physica A 387, 915 (2008).

[8] A. Porporato, L. Ridolfi, Multivariate nonlinear prediction of river flows, J. Hydrol. 248 (2001) 109–122.

[9] R. Stoop, N. Stoop, L. Bunimovich, Complexity of dynamics as variability of predictability. J Stat. Phys. 114 (2004) 1127–1137.

[10] Y.M. Otache, M.A. Sadeeq, I.E. Ahaneku, ARMA modelling of Benue River flow dynamics: Comparative study of PAR model, Open J. Modern Hydrol: 1 (2011) 1-9.

[11] M. Li and P. Vitanyi, An Introduction to Kolmogorov Complexity and its Applications, Second Edition, Springer Verlag, Berlin, 1997, pages 1-188.

[12] A. Lempel, J. Ziv, On the complexity of finite sequences. IEEE Trans. Inform. Theory 22 (1976) 75–81.

[13] I. Krasovskaia, Entropy-based grouping of river flow regimes, J. Hydrol. 202 (1997) 173-191.

[14] V. P. Singth, The use of entropy in hydrology and water resources, Hydrol. Process. Vol 11 (1997) 587-626.

[15] S.M. Pincus, Approximate entropy (ApEn) as a complexity measure, Chaos 5(1):110–7 (1995).

[16] J.S. Richman, J.R. Moorman, Physiological time-series analysis using approximate entropy and sample entropy, Am. J. Physiol. Heart Circ. Physiol. 278 (2000) H2039–H2049.

[17] C.M. Chou, Applying multiscale entropy to the complexity analysis of rainfall-runoff relationships, Entropy: 14 (2012) 945-957.

[18] C. Bandt, B. Pompe, Permutation entropy: a natural complexity measure for time series, Phys. Rev. Lett. 88:174102 (2002).

[19] W.J. Pei, Z.-Y. He, L.X. Yang, S.S. Hull, J.Y. Cheung, A statistical complexity measure and its applications to the analysis of heart rate variability, Acoustics, Speech, and Signal Processing, 2000. ICASSP '00. Proceedings. 2000 IEEE International Conference on Acoustics, Speech, and Signal Processing Proceedings vol.1 (2000) 185-188.

[20] R. Lopez-Ruiz, H.L. Mancini, X. Calbet, A statistical measure of complexity, Phys. Lett. A 209 (1995) 321-326.

[21] P. W. Lamberti, M. T. Martin, A. Plastino, O. A. Rosso, Intensive entropic non-triviality measure, Physica A 334 (2004) 119-131.

[22] O. A. Rosso, L. Zunino, D. G. Perez, A. Figliola, H.A. Larrondo, M. Garavaglia, M. T. Martin, A. Plastino, Extracting features of Gaussian self-similar stochastic processes via the Bandt-Pompe approach, Phys. Rev. E 76, 061114 (2007).

[23] D. P. Feldman, C. S. McTague, and J. P. Crutchfield, The organization of intrinsic computation: Complexity-entropy diagrams and the diversity of natural information processing, Chaos 18, 043106 (2008).

[24] X.S. Zhang, R.J. Roy, E.W. Jensen, IEEE Trans. Biomed. Eng. 48: 424 (2001).

[25] N. Radhakrishnan, J.D. Wilson, P.C. Loizou, An Alternative Partitioning Technique to Quantify the Regularity of Complex Time Series. Int. J. Bifurc. Chaos 10 (2000) 1773-1779.

[26] M. Small, Applied Nonlinear Time Series Analysis: Applications in Physics, Physiology and Finance, World Scientific, Singapore, 2005.

[27] R. Ferenets, T. Lipping, A. Anier, Comparison of entropy and complexity measures for the assessment of depth of sedation, IEEE Trans. Biomed. Eng. 53:1067 (2006).
[28] Thai Q, http://www.mathworks.com/matlabcentral/fileexchange/38211-calclzcomplexity (2012)

[29] M.B. Kennel, R. Brown, H.D.I. Abarbanel, Determining embedding dimension for phase-space reconstruction using geometrical construction, Phys. Rev. A 45: 3403 (1992).

[30] D.E. Lake, J.S. Richman, M.P. Griffin, J.R. Moorman, Sample entropy analysis of neonatal heart rate variability, Am. J. Physiol. Heart C. 283: R789 (2002).

[31] S.M. Pincus, Approximate entropy as a measure of system complexity, Natl. Acad. Sci. USA 88:2297 (1991).!

[32] E. Hadžić, N. Drešković, Climate change impact on water river flow: A case study for Sarajevo Valley (Bosnia And Herzegovina), In: Mihailović DT (ed) Essays on Fundamental and Applied Environmental Topics, Nova Science Publishers, New York, 2012, pp. 307-332.

[33] F.F. Ferreira, G. Francisco, B.S. Machado, P. Murugnandam, Time series analysis for minority game simulations of financial markets, Physica A 321 (2003) 619–632.

[34] V. Djurdjevic, B. Rajkovic, Verification of a coupled atmosphere-ocean model using satellite observations over the Adriatic Sea, Ann. Geophys. 26 (2008) 1935-1954.

[35] J.M. Yentes , N. Hunt, K.K. Schmid, J.P. Kaipust, D. McGrath, N. Stergiou, The Appropriate use of approximate entropy and sample entropy with short data sets, Ann. Biomed. Eng. (2012).
http://link.springer.com/content/pdf/10.1007%2Fs10439-012-0668-3. 3 January 2013

[36] L.D. Frank, J.F. Sallis, T.L. Conway, J.E. Chapman, B.E. Saelens, W. Bachman, Many pathways from land use to health: Associations between neighborhood walkability and active transportation, body mass index, and air quality, J. Am. Plann. Assoc. 72(1) (2006) 75–87.

[37] J. Hu, J. Gao, J.C. Principe, Analysis of biomedical signals by the Lempel-Ziv complexity: the effect of finite data size, IEEE T. Bio-Eng: 53:2606-9 (2006).

[38] N.D. Gordon, T.A. McMahon, B.L. Finlayson, C.J. Gipel, Stream hydrology: an introduction for ecologists, Wiley, New York, 2004.

[39] H.G. Orr, P.A. Carling, Hydro-climatic and land use changes in the river Lune catchment, North West England, implications for catchment management, River Res. Appl. 22 (2006) 239–255.